\documentclass[aps,twocolumn,showpacs,preprintnumbers,amsmath,amssymb,floatfix]{revtex4}
\usepackage{graphicx}
\usepackage{epsfig}
\begin{document}
\def\be{\begin{equation}}
\def\ee{\end{equation}}
\def\bearr{\begin{eqnarray}}
\def\eearr{\end{eqnarray}}
\def\tc{$T_c~$}
\def\bis2{$\rm BiS_2$~}
\def\spx{$\rm 6p_x$~}
\def\spy{$\rm 6p_y$~}
\def\laob{$\rm LaOBiS_2$~}
\def\laofh{$\rm LaO_{0,5}F_{0.5} BiS_2$~}
\def\laofx{$\rm LaO_{1-x}F_{x} BiS_2$~}
\def\half{$\rm \frac{1}{2}$~}

\title{Theory of Ultra Low Tc Superconductivity in Bismuth: Tip of an Iceberg ?}

\author{G. Baskaran}

\affiliation
{The Institute of Mathematical Sciences, C.I.T. Campus, Chennai 600 113, India \&\\
Perimeter Institute for Theoretical Physics, Waterloo, ON N2L 2Y5, Canada}

\begin{abstract}
Superconductivity with an ultra low Tc $\sim$ 0.5 mK was discovered recently in bismuth, a semimetal. To develop a model and scenario for Bi we begin with a cubic reference lattice, close to A7 (dimerized cubic) structure of Bi. Three valence electrons hop among 6p$_x$, 6p$_y$ and 6p$_z$ orbitals and form \textit{quasi one dimensional chains at half filling}. An interesting interplay follows: i) Mott localization tendency in the chains, ii) metallization by interchain hopping and iii) lattice dimerization by electron-phonon coupling. In our proposal, a potential high Tc superconductivity from RVB mechanism is lost in the game. However some superconducting fluctuations survive. Tiny fermi pockets seen in Bi are viewed as remnant \textit{evanescent Bogoliubov quasi particles} in an anomalous normal state. Multi band character admits possibility of PT violating \textit{chiral singlet superconductivity}. Bi has a strong spin orbit coupling; Kramers theorem protects our proposal for the bulk by replacing real spin by Kramer pair. Control of chain dimerization might resurrect high Tc superconductivity in Bi, Sb and As.
\end{abstract}
\maketitle

\section{Introduction}

Semimetals that straddle insulators and metals are fascinating systems. Elemental semimetals, graphene, $\alpha$-Sn, As, Sb and Bi are special, partly because of their structural simplicity. Two dimensional graphene, a relatively recent entry has created a great excitement in the scientific community. Bi, on the other hand, has been investigated \cite{BiRussianReview} for more than a century, from basic physics point of view. Phenomena such as, diamagnetism, Nernst effect, de Haas van Alphen quantum oscillation etc., were discovered \cite{Fukuyama} for the first time in Bi, inspite of its ultra low carrier density $\sim$ 3 x 10$^{17}$ /cm$^3$. In recent times, anomalies in quantum oscillations, quantum Hall phenomena \cite{BehniaKopelevich,CavaOng,QHallSurface}, Nernst effect \cite{NernstEffectBehnia}, optics \cite{vanDerMarel}, NMR \cite{CavaNMR}, pressure effects \cite{BiPressure}, laser induced structural changes \cite{BiStructureLaser} and topological electronic phases \cite{MurakamiKaneCavaTopology,Fukuyama} on surfaces of Bi have been catching the attention of physics community. 

Solid Bi does not exhibit superconductivity. It is not a surprise, as Bi has about one free carrier per 10,000 Bi atoms. However, Bi behaves like a fermi liquid with small fermi pockets, leading to an expectation that it might superconduct at sufficiently low temperatures. On the other hand, small fermi energy and non-adiabatic effects make phonon mediated superconductivity doubtful \cite{BiTIFR}. 

It is in this background discovery of type I superconductivity in ultra pure Bi crystal by the TIFR group \cite{BiTIFR}, with an ultra low Tc $ \sim $ 0.5 mK, has come as a surprise. An element of surprise (sic).
In terms of low carrier density superconductors, closest to Bi is lightly doped SrTiO$_3$. In a recently studied doped SrTiO$_3$ \cite{STO,STOTheory} carrier density is $\sim$ 50 times larger and Tc is in the scale of 100 mK. Further, Tc in Bi is close to the record low (finite) Tc $\sim$ 0.3 mK seen in Rhodium \cite{Rhodium}, a metal having a high carrier density $\sim 10^{22}/cm^3$. 

Interestingly, pure Bi seems to be at the \textit{verge of superconductivity, even in the Kelvin scale}. It readily superconducts when perturbed \cite{PerturbedBiSupCond}. Pressurized crystalline Bi, disordered thin films,  amorphous Bi, interfaces, nano wires etc., exhibit superconductivity with Tc's in the range of 6 - 8 K. Even though Tc is small, it is a four order of magnitude jump from 0.5 mK ! There has been also reports \cite{BiSupCond36K} of interfacial superconductivity in bicrystals of Bi and Bi$_{1-x}$ Sb$_x$ alloys, with a Tc onset as high as 36 K.  

Bi has been theoretically investigated extensively, as a band semi metal \cite{MHCohenWolf} and a satisfactory and quantitative understanding of its low energy electrical and magnetic properties is believed to exist. In these theoretical attempts a large spin orbit coupling in Bi plays an important role \cite{BiBandStructure}. Coulomb interactions in Bi has been studied in the past while discussing possibility of exciton condensation and Wigner crystallization \cite{BiRussianReview,BiExcitonCond}. A very recent work by Koley, Laad and Taraphder \cite{Koley} that followed heels of experimental discovery, offers an exciton fluctuation mediated mechanism of superconductivity in Bi. In an attempt to understand normal state properties of Bi quantitatively, Craco and Leoni \cite{BiCracoLeoni} have emphasized importance of using a moderate Hubbard U in the 6p orbitals of Bi. 

Briefly, in our theory a potential high Tc superconductivity in crystalline Bi has been pushed down to an ultra low, milli Kelvin temperature scale. Superconducting fluctuations however survive even at room temperatures and makes normal state anomalous. Our theory suggests in a natural fashion ways to resurrect high Tc superconductivity in Bi and its neighbours Sb and As, in the same column in the periodic table.

To motivate our model and scenario, we begin with a known fact \cite{PeierlsBook,RoldHoffmann} that weakly coupled quasi one dimensional tight binding bands make three dimensional bands in the A7 structure of Bi, Sb and As. Three sets of mutually perpendicular tight binding chains at half filling emerge from a \textit{cubic lattice based orbital organization of three valence electrons in the 6p$_x$, 6p$_{y}$ and 6p$_z$ orbitals} of Bi. The quasi one dimensionality, together with an available moderate Hubbard U in 6p orbitals of Bi leads to remarkable possibilities and an interesting interplay. This is because even a small Hubbard U in one dimensional Hubbard chain is relevant; for example, at half filling any finite repulsive U leads to Mott localization and opens a charge gap, and maintains a zero spin gap \cite{LiebWu}. 

In our theory, there is an interesting interplay of three competing phenomena and tendencies: i) Mott localization in the chains, ii) metallization by interchain hopping and iii) valence bond trapping and chain dimerization by electron-lattice coupling. It leads to a rich scenario and possibilities that is hidden in semimetallic solid Bi.

Our paper is organized as follows. We present a model that manifestly brings out a hidden Mott physics in the form of weakly coupled Mott Hubbard chains at half filling. Then we discuss, using known results, how small interchain electron hopping (small relative to intrachain hopping) causes a Mott insulator to metal transition, but keeps the metal close to the Mott insulator boundary (figure 2). This is followed by discussion of a potential RVB physics and high Tc superconductivity in this correlated metallic state. 

Next section discusses an inevitable electron lattice coupling that traps valence bonds by chain dimerization, leading to major consequences: i) disappearance of a potential high Tc superconductivity, ii) surviving superconducting fluctuations in an anomalous normal state and iii) remnant \textit{evanescent Bogoliubov quasi particles} in the form of small electron and hole fermi pockets in the BZ.

Ways by which we could control lattice dimerization and resurrect high Tc superconductivity in Bi as well as in isostructural and isoelectronic Sb and As are discussed next. At the end  we make some comments about normal state anomalies in Bi, in the light of our proposal.

\begin{figure}
\includegraphics[width=0.3\textwidth]{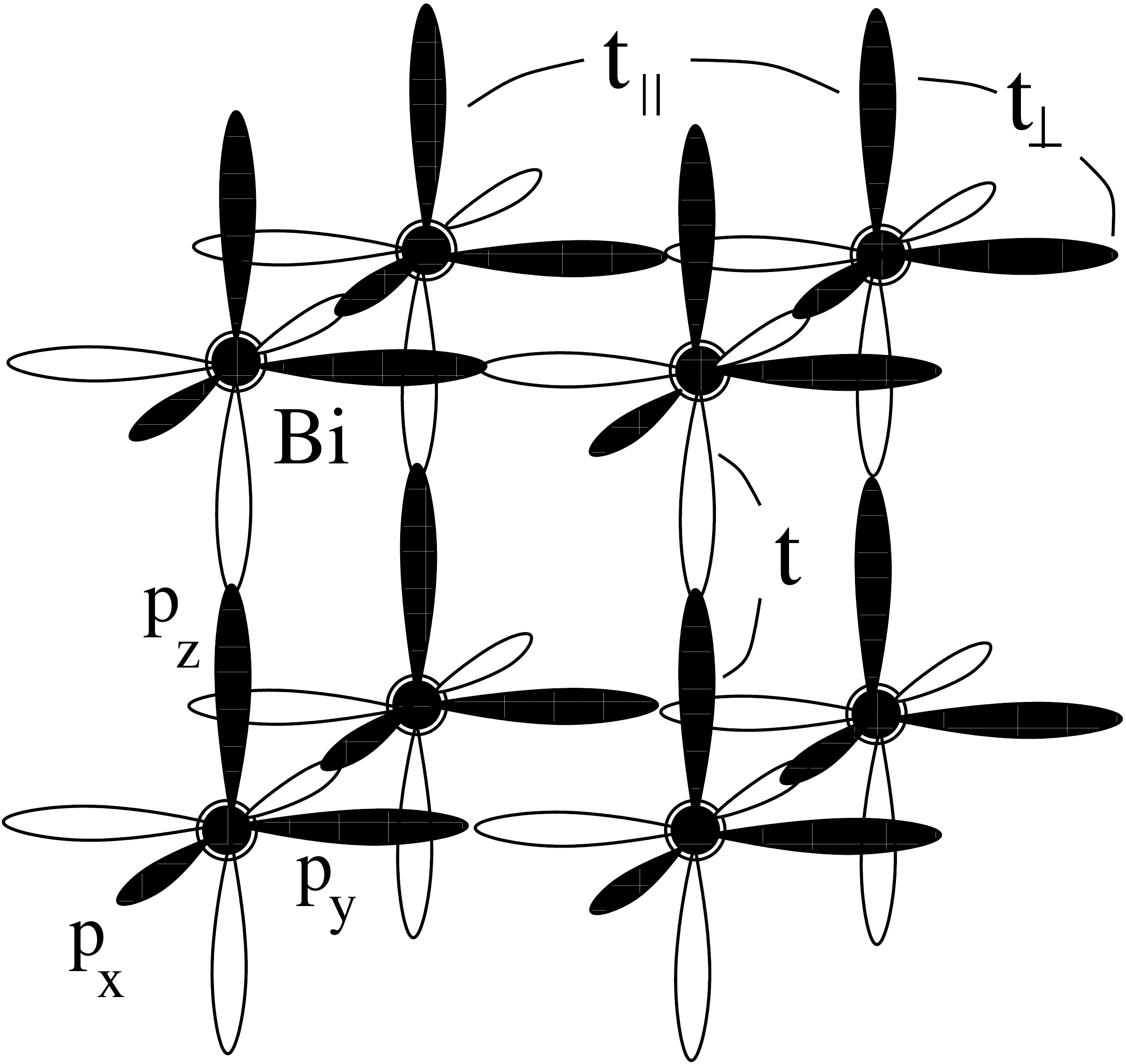}
\caption{\textbf{Orbital organization} of 6p orbitals into mutually perpendicular (half filled band) chains
in a reference cubic lattice for Bismuth. From known band structure results \cite{BiBandStructure}, intrachain hopping t $\approx$ 1.85 eV, t$_{||} \approx$ 0.4 eV and t$_{\perp} \approx$ 0.2 eV.} \label{Fugure 1}
\end{figure}

\section{Model and a Reference State} 

Bi has a distorted cubic A7 structure \cite{BiRussianReview}. Electronic configuration of Bi atom is [Xe] [4f${^{14}}$ 5d$^{10}$ 6s$^2$] 6p$^3$. Three valence electrons in the half filled 6p$^3$ shell essentially determine low energy physics of solid Bi. Filled 6s band lies nearly 10 eV below the fermi level \cite{BiBandStructure}. It has been well recognized that 6p$_x$, 6p$_y$ and 6p$_z$ orbitals strongly hybridize along respective x, y and z directions and form weakly coupled one dimensional bands that are half filled. In the nearly cubic structure, interchain hopping is relatively weak.

To build on known non-interacting tight binding model, we start with \textit{an undistorted cubic lattice of Bi atoms} (Figure 1) \textit{as a reference solid} \cite{RoldHoffmann}. Most importantly, we recognize that coulomb interaction in the quasi one dimensional tight binding model can't be ignored.

Our model Hamiltonian has four parts:
\bearr 
H &=& H_{\rm c} + H_{\rm ic} + H_{\rm lrc} + H_{ep} 
\eearr

First term H$_{\rm c}$ is the Hubbard Hamiltonian of the chains. Second term H$_{\rm ic}$ contains interchain hopping. Third term, H$_{lrc}$ is the long range Coulomb interaction term, beyond on site Hubbard U. Last term, H$_{ep}$ is the sum of phonon and electron-phonon coupling Hamiltonians.

Coupled chain Hamiltonians containing p$_x$, p$_y$ and p$_z$ orbitals (generalized three orbital Hubbard models) in square \cite{KivelsonTelluride} and cubic lattices have been investigated, including exact results for itinerant ferromagnetism \cite{LiLiebWu} for some choice of parameters. In our problem we are in a different region of parameter space and at half filling, where singlets dominate.

\textbf{Decoupled One Dimensional Chains as a Reference System:}
Chain Hamiltonian, which provides us a convenient reference state is:
\bearr
H_{\rm c} &=& -t \sum_{\textbf{i}\alpha \sigma} (c^\dagger_{\textbf{i}\alpha \sigma}c^{}_{\textbf{i}
+\textbf{a}_{\alpha}\alpha\sigma} + h.c.) + U \sum_{\textbf{\textbf{i}},\alpha} n_{\textbf{i}\alpha \uparrow} n_{\textbf{i}\alpha \downarrow}~~~~
\eearr

Cubic lattice sites are denoted by \textbf{i}; $\textbf{a}_{\alpha}$'s are nearest neighbour lattice vectors along  $\alpha$ = x,y and z directions. The c operators are electron operators; $\sigma$ is the spin index and $\alpha$ = x,y and z are orbital indices. From band theory \cite{BiBandStructure} results we find that the intrachain nearest neighbour hopping term t $\sim$ 1.85 eV. Hubbard U, estimated in reference \cite{BiCracoLeoni}, required for a quantitative understanding normal state and high energy properties of Bi is U $\sim$ 5 to 8 eV.

We first focus on the half filled band Hubbard chain Hamiltonian H$_{\rm c}$, which provides us a convenient reference phase. We use Lieb-Wu exact solution \cite{LiebWu} to understand the above decoupled 3d network of one dimensional nearest neighbour repulsive Hubbard chain Hamiltonian. From Lieb-Wu solution we find that {Mott-Hubbard gap is $\sim $ 1 to 1.5 eV, for a chain, when t $\approx$ 1.85 eV and U $\approx$ 5 to 8 eV. 

\begin{figure*}
\includegraphics[width=0.8\textwidth]{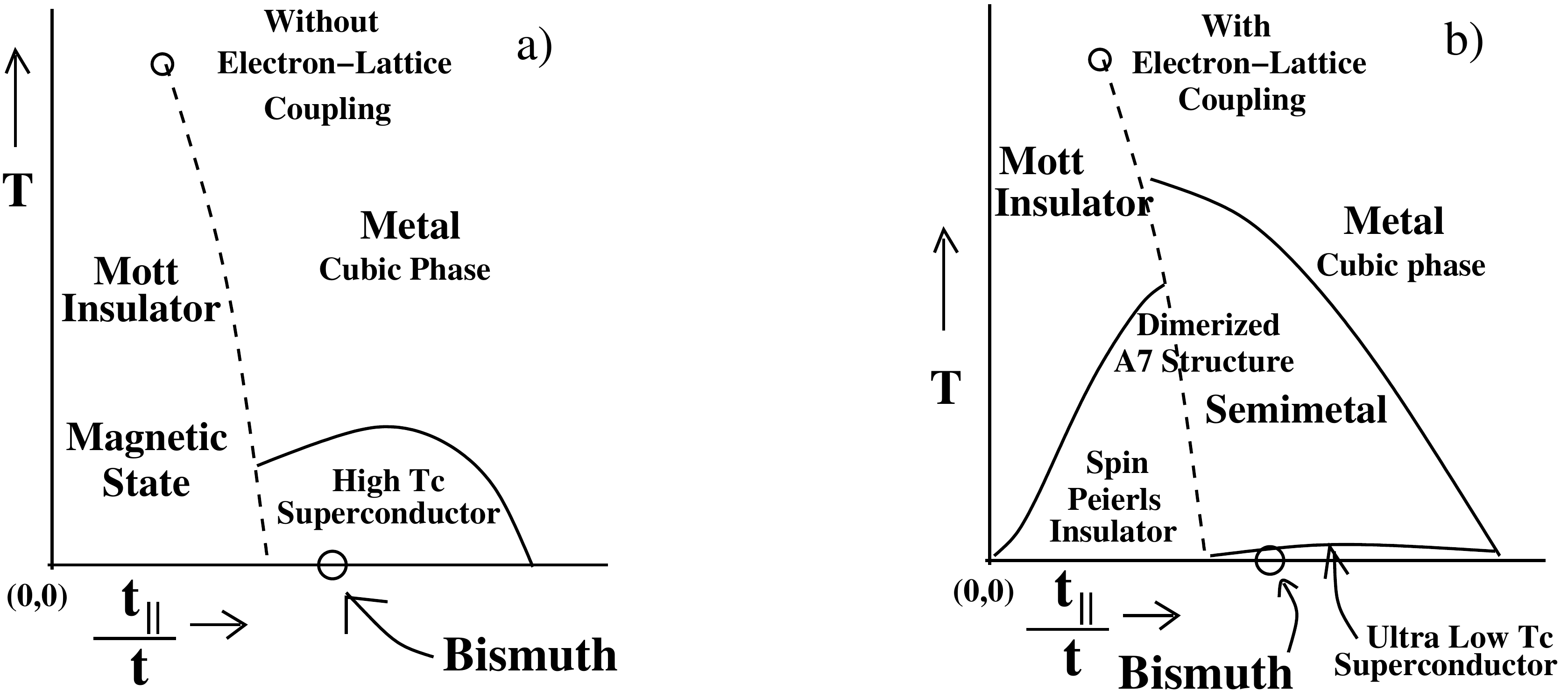}
\caption{
\textbf{Schematic Phase Diagram} for model Hamiltonian (equation 1), in the temperature and interchain hopping
$\frac{t_\perp}{t}$ plane; here t is the intrachain hopping. a) Reference Cubic Phase in the absence of electron-lattice coupling: A first order phase transition line, ending at a critical point, separates Mott insulator and metal. Location of Bi solid in the t$_{\perp}$-axis is marked. b) Presence of electron-lattice coupling: Spin Peierls Insulator and semimetallic phase in A7 dimerized cubic structure, high temperature cubic metallic phase and ultralow superconductivity are shown.
} \label{Figure 2}
\end{figure*}

This is a key result: \textit{a robust Mott Hubbard gap in the range of 1 to 1.5 eV exists, in the decoupled chains of the reference state} we start with. This Mott insulating Hubbard chain has a spin liquid ground state- It contains enhanced nearest neighbour spin singlet pairing correlations and supports gapless spinon excitations and gapful charge excitations. When we consider the three dimensional network of chains, spinon excitations form three sets of pseudo fermi surfaces (sheets) in the three dimensional BZ, lying parallel to xy, yz and zx planes.

Since we are dealing with Mott insulating chain with a reasonable charge gap, low energy spin dynamics is well approximated by three sets ($\alpha$ = x,y and z) of Heisenberg chain Hamiltonians:

\be 
H_{\rm c} \approx  \frac{J}{2}\sum_{\textbf{i}\alpha} (\textbf{S}_{\textbf{i}\alpha} \cdot \textbf{S}_{\textbf{i}
+\textbf{a}_{\alpha}\alpha} - \frac{1}{4})
\ee

The effective \textit{superexchange} (kinetic exchange)  is traditionally approximated by J = $\frac{U}{2} [1 + \frac{16t^2}{U}]^{\frac{1}{2}} - \frac{U}{2}$, the energy difference between excited triplet and singlet ground state of two electrons on neighbouring p-orbitals, in a given chain. It is a measure of spin singlet correlations, in the background of fluctuating charges. In the large U limit we get the standard superexchange term, J $\approx \frac{4t^2}{U}$. We find that the effective exchange parameter for Bi is J $\approx$ 1 eV; it is large and comparable to the Mott Hubbard charge gap within a chain.

\textbf{Interchain Hopping and Transition from Mott Insulator to Correlated Metallic State:} Repeated electron electron scattering in one dimensional tight binding chain builds a Mott gap for arbitrarily small value of U, at half filling. However, interchain hopping and residual long range interaction could close Mott Hubbard gap and cause metallization. The transition takes place when the \textit{energy gain from three dimensional delocalization by interchain hopping becomes comparable to Mott Hubbard gap}. In real systems Mott transition is a first order transition (rather than 2nd order as given by Hubbard model), inview of the long range part of the coulomb interaction.

From theory point of view, before the first order transition we have weakly coupled three dimensional network of spin-half Heisenberg chains. This magnetic state could support antiferromagnetic order and three dimensional spin liquid ground states, depending on the parameters (figure 2a).  

Evolution of the fermi surface on the metallic side of the Mott transition, as a function of interchain hopping \cite{InterchainCouplingMetallization} or pressure is interesting. Correlated metallic state close to the Mott transition boundary might support low energy and gapped spin-1 bound state branch in some parts of the BZ. That is, fluctuating Mott localization in the metallic chains is likely to support gapful spin-1 excitations, in regions where spinon fermi surface  existed prior to metallization.

Now we consider interchain hopping terms (figure 1) present in the Hamiltonian H$_{\rm ic}$ for Bi. Nearest neighbour interchain hopping between same type of chains, for example x chains is t$_\parallel$.
Coupled x chains alone form an anisotropic 3 dimensional tight binding lattice. Similarly, y and z chains form their own anisotropic 3 dimensional lattices. These three systems are coupled by a smaller interchain hopping t$_{\perp}$. From existing band theory \cite{BiBandStructure} we find that the largest interchain hopping is t$_{\parallel} \sim $ 0.4 to 0.6 eV, This is less than one third of the nearest neighbour intra chain hopping t $\approx$ 1.8 to 2  eV. The next leading interchain hopping is t$_{\perp} \approx$ 0.2 eV.

Major perturbation arising from interchain hopping can gain delocalization energy $\sim z t_{\parallel} \approx$ 1.5 eV, where z = 4 is the number of nearest parallel chains to a given chain. Since this energy is close to our estimated charge gap $\sim$ 1 to 1.5 eV in the Mott chain, it is strongly suggestive that the cubic reference system for Bi is in a metallic state close to the Mott transition boundary (Figure 2). 

\textbf{Potential High Tc RVB Superconductivity:}
In the resonating valence bond (RVB) theory \cite{RVBTheory}, singlet pair correlations arising from superexchange in a doped Mott insulator is a key requirement for high Tc superconductivity. Where is superexchange in the correlated half filled band metallic state ? 

The conducting state stabilised by small interchain hopping and coulomb interaction gains delocalization energy and Madelung energy. This correlated state has been viewed by the present author \cite{GBOrganics} as a \textit{self doped Mott insulator} in the following sense. Even while maintaining superexchange and Mottness locally, the system spontaneously creates a small and equal density of holons (an empty p orbital) and doublons (a doubly occupied p orbital) and sustains it.  Density of self doping is determined self consistently by coulomb and band parameters. 

Survival of superexchange and presence of self doping are seen in the optical conductivity of organic conductors, close to Mott transition point; the former as a Mott Hubbard gap and the later as a small Drude peak \cite{GBOrganics,OrganicsOptics}. 

Important scales in this correlated metallic states are t, the hopping matrix element, J the surviving superexchange, and n the density of self doping. This is contained in an effective model called 2 species tJ model \cite{GBOrganics}:

\bearr
H_{tJ}  = &-& \sum_{\textbf{i} \textbf{j}\alpha \beta }t^{\alpha\beta}_{\textbf{i j}}~~{\hat P}_{\rm hd}~(c^\dagger_{\textbf{i}\alpha\sigma}c^{}_{\textbf{j}\beta\sigma} + H.c. ){\hat P}_{\rm hd} + \nonumber \\
& + & \frac{J}{2} \sum_{\textbf{i}\alpha} ( {\bf S}_{\textbf{i}\alpha} \cdot {\bf S}_{\textbf{i} + \textbf{a}_{\alpha} \alpha} - 
\frac{1}{4} n^{}_{\textbf{i}\alpha} n^{}_{\textbf{i} + \textbf{a}_{\alpha} \alpha} ) 
\eearr

In the above Hamiltonian, projection operator ${\hat P}_{\rm hd}$ ensure that as holons and doublons hop they don’t' annihilate each other. This ensures that number of doublons and number of holons, which are equal, are individually conserved. In the background of this conserved number of dynamic doublons and holons we have singly occupied sites containing spins. We have already discussed interchain hopping matrix elements $t^{\alpha\beta}_{\textbf{i j}}$: they are  t$_{\parallel} \sim$ 0.4 to 0.6 eV and t$_{\perp} \sim 0.2 $ eV. 

In the above Hamiltonian we have retained the largest of J, between nearest neighbour sites within a given chain. It is straight forward to perform RVB mean field theory for the above Hamiltonian directly or using slave-particle formalism. We will not go into details but point out that we get very high Tc superconductivity within mean field theory. Primary reason is the large t and J that we have identified. However, strong one dimensional fluctuations, very small self doping etc. are likely to reduce superconducting Tc further. 
 
There is a heuristic way to estimate superconducting Tc within RVB mechanism. As temperature is lowered spinons get paired and charged valence bonds dominate. At a particular temperature charged valence bonds (whose density is the same as holon and doublon density) undergo Bose Einstein condensation \cite{KivelsonShortRangeRVB}, resulting in superconductivity. We can use the Bose Einstein condensation formula and estimate superconducting Tc:

\be
k_B Tc \approx 3.3125 ~ \frac{\hbar^2 n^{\frac{2}{3}}}{m^* k_{\rm B}}
\ee

In this simple expression for Tc there are two unknown parameters, a mean effective mass m$^*$ and n, density of self doped carriers. Unfortunately these parameters are known experimentally only in the dimerized, distorted cubic, A7 structure. If we make a reasonable guess of the two parameters in the reference undistorted cubic phase, we get superconducting Tc in the scale of 100 K.

Now we will discuss a lurking danger for high Tc superconductivity from a competing valence bond order.

\textbf{Valence Bond Order and Collapse of Superconducting Tc:}
We begin by discussing the effect of electron-phonon interaction on the reference Mott insulating chains. Then discuss its drastic effect on quasi three dimensional correlated metallic state, which support strong pairing correlations and high Tc superconductivity.

In a single band free fermion chain on a deformable lattice a 2k$_F$ instability opens a gap at the fermi level. This is the well known Peierls instability \cite{PeierlsBook}. It arises from a singular response of one dimensional fermi gas to perturbations at wave vector 2k$_F$. After the Peierls distortion we have a band insulator; a finite gap exists for any type of electronic excitation.

Same dimerization phenomenon in a Mott-Hubbard chain at half filling, has interesting extra physics because of presence of spin-charge decoupling. In the undistorted Mott Hubbard chain we have two type of low energy excitations. A gapless spinon excitation branch and a gapful charge excitation branch. Dimerization opens up gap in the spin spectrum and makes an extra contribution to the already existing large charge gap. 

Even after dimerization spin charge decoupling continues in the following fashion. A gapful low energy spin-1 branch (bound state of two spinons) lies below the charge gap. Similar physics exists in polyacetylene, a one dimensional tight binding system. Here the important role of electron correlation effects \cite{RamaseshaSoos}, on top of electron-phonon coupling \cite{SuSchriefferHeeger} has been well recognized. Dimerization arising from electron-phonon interaction in a Mott insulating chain is called spin-Peierls instability; because the soft spin degree of freedom are mostly getting affected by dimerization (Figure 2b).

The situation in the quasi one dimensional correlated metallic state is somewhat complex. Major complication is presence of a long range superconducting order and strong superconducting pairing correlation
and valence bond resonance that supports it. It is obvious that valence bond localization and lattice dimerization continues in the quasi one dimensional case as well. This is manifest when we look at A7 structure of Bi.

Valence bond order takes place in a cooperative fashion in all x,y and z - chains, leading to the observed distortion of the cubic structure to A7 structure. Bond charge repulsion arising from the three double bonds emanating from every Bi atom leads to the small ($\sim 3^0$) rhombohedral angular distortion. From experimental point of view A7 structure does not change to a cubic structure till the melting point. That is, free energy gain from valence bond order is high. Valence bond order is a strong competitor to superconductivity - no wonder that superconductivity is suppressed (figure 2b) to milli Kelvin scale in Bi.

A closer look reveals interesting hidden physics in the form of superconducting fluctuations till room temperature scales in crystalline Bi.

\textbf{Survival of Superconducting Fluctuations and remnant Evanescent Bogoliubov Quasi particles in the Normal State:} Before electron lattice coupling intervened, metallic state has a strong local pairing and a finite zero momentum Cooper pair condensate fraction, a high Tc superconducting state. In the superconducting state low energy excitations are Bogoliubov quasi particles; in our mean field theory they have a finite (superconducting) gap. 

We imagine turning on electron-phonon interaction adiabatically in this superconducting state. Strength of zero momentum condensate will continuously decrease, as competing valence bond order and chain dimerization grow. We interpret development of valence bond order as build-up of a commensurate finite momentum pair condensate at the reciprocal lattice vector of the A7 structure, at the expense of zero momentum condensate. At the same time superconducting gap decreases and gives way for valence bond order. During this evolution, superconducting gap is likely to close at some regions in k-space.

We interpret what is seen in real Bi as a remnant of superconducting fluctuations after superconducting Tc has crashed to 0.5 mK. In this case it is natural to view existing electron and hole pockets at L and T points in the BZ as remnant \textit{evanescent Bogoliubov quasi particles} in the normal state, in the background of superconducting fluctuations. 

By evanescent Bogoliubov quasi particle we mean the following. A real Bogoliubov quasi particle is a coherent superposition of an electron and hole state. It reflects presence of a finite zero momentum condensate of Cooper pairs present in the vacuum. We envisage that in the presence of a fluctuating condensate the electron and hole quasi particles in Bi will have transient Bogoliubov quasi particle character, through Andreev reflection by a strongly fluctuating local phase order. 

An electron or hole quasi particle in a standard band insulator does not have superconducting fluctuation in the vacuum. In this sense the vacuum that supports the electron and hole quasi particles and the superconducting pairing fluctuations in Bi are different from a standard band insulator vacuum.  

Strong pairing correlation in Bi is a reservoir which represents resonating valence bonds. Since the effective superexchange J is high, valence bonds are highly quantum. \textit{Consequently the valence bond solid is a quantum solid, a crystal of paired electrons with strong pair fluctuations}. 

Normal state in Bi has some similarity to the pseudogap phase of cuprates, where valence bonds get trapped by lattice distortions and form ordered stripes (valence bond order), at the expense of superconductivity. In cuprates we have external doping; in bismuth we have self doping. In the 1/8-th commensurate doped 
La$_{2-x}$Ba$_x$CuO$_4$ cuprate we have a commensurate valence bond order and a nearly vanishing superconducting Tc.

\textbf{Recovery of Superconductivity by Quantum Melting of Valence Bond Solid:}
Is there a way of quantum melting the competing valence bond order and resurrecting high Tc superconductivity ?
At least two ways seem possible. One is external doping and destabilization of valence bond order. Alloys such as Bi$_{1-x}$A  (A = Pb, Sn, Sb, As, P, Tl, Se, Te, ..) formed from neighbouring elements in the periodic table hold some promise. If alloying converts the A7 structure to a simple cubic structure, there exists prospects for superconductivity, with a Tc much higher than 0.5 mK ! Atomic radius and suitable quantum chemistry issues 
needs to be taken care of in this approach.

In one of the experiments Bi$_{1-x}$Sb$_x$ shows signals of superconductivity with a Tc onset as large as 36 K \cite{BiSupCond36K}. It is likely that in the bicrystal interface, because of lattice mismatch and strain, interface region acquires local cubic character. It results is quantum melting of valence bond order and recovery of RVB state in the interface region. This could enhance superconducting Tc.

Use of hydrostatic or uniaxial pressure is another route. Early works and a very recent work \cite{BiPressure} points out that in single crystal Bi pressure brings in a series of structural changes and also makes superconducting Tc as high as 6 to 8 K. 

On closer inspection we find that pressure induced new structures have decreasing valence bond ordering in different fashion. Experimentally a simple cubic phase does not get stabilized at finite pressure in Bi however. The closest is a cubic phase with monoclinic distortion, which is superconducting with a Tc of about 6 K. An interesting structure called \textit{host-guest phase}, occurring at higher pressure also has finite Tc superconductivity. In this structure superconductivity is likely to arise from weakly coupled guest chains of Bi embedded in the covalent insulating network of host three dimensional Bi lattice. We hope to discuss these structures in a future publication.

As we mentioned earlier As and Sb have the same A7 structure as Bi and are isoelectronic. The physics we have discussed is relevant there. It is interesting that As has a cubic structure at a range of moderate pressure, in addition to other structures. Nature of low temperature phase in the cubic structure is worth investigating further. Our proposal predicts recovery of valence bond resonance and resurrection of high Tc superconductivity in the undistorted cubic structure of As.

\textbf{Possibility of PT Violating Chiral Superconductivity:}
In the theory we have presented so far, each one of the x,y and z - chain system becomes an anisotropic three dimensional system. Every one of them is a three dimensional self doped Mott insulator. They could individually support high Tc superconductivity based on RVB mechanism. Residual interaction among these self doped Mott insulating systems could bring in new possibilities for symmetry of superconducting order parameter.

It is known that repulsive coulomb interaction can cause zero momentum electron pair tunnelling (scattering) between the 3 sets of bands. Repulsive pair scattering will frustrate relative phases of the s-wave order parameters in the three systems. In the context of multiband superconductivity this problem has been studied in the past. In the case of 2 bands with repulsive pair scattering, Kondo \cite{Kondo} finds that phases of the order parameter change sign between two bands. If we have three bands related by symmetry, repulsive pair scattering could lead to \cite{Tesanovic} two degenerate \textit{PT violating chiral superconducting states} having s-wave singlet order parameter and a relative phase difference between three bands: (0, $\frac{2\pi}{3}$, $\frac{4\pi}{3}$) or (0, - $\frac{2\pi}{3}$, - $\frac{4\pi}{3}$).

What happens after a high Tc superconductivity crashes, once valence bond order and lattice dimerization sets in ? We find that the situation is somewhat similar for pairing in the electron and hole pockets. That is, repulsive inter band scattering among pockets at L and T points favours a PT violating ultra low Tc chiral superconducting state.

\textbf{Normal State Anomalies:} 
In our theory Bi is a kind of supersolid of Cooper pairs. A quantum crystal of paired electrons coexists with  phase fluctuating zero momentum Cooper pairs. It is likely that phase fluctuations are organized into vortices and fluctuating circulating diamagnetic currents, reminiscent of Anderson-Ong's theory \cite{AndersonOng} of vortex liquid for the pseudo gap phase of cuprates. In this sense normal state of Bi in our theory, following Anderson's analogy \cite{PWASupSolidCuprate} for pseudogap phase of cuprates, is similar to supersolid phase in $^4$He. New aspect here is a small density of electron and hole carriers present in this rich background, behaving as remnant evanescent Bogoliubov quasi particle. 

Some of the striking anomalies in the normal state of Bi are long mean free paths, large Nernst and diamagnetism signals, a rich magnetic field induced quantum oscillation phenomena etc. There has been serious study in the past to explain these anomalies using one electron physics, based on the peculiar band structure of Bi and the strong spin-orbit coupling. Our thesis is that pairing correlation is very basic and unavoidable in solid Bi. Such hidden electron pairing correlations, not contained in a standard band insulator description, could make substantial contribution to the above anomalies.

One of the less known and little understood anomalies in Bi is a large scattering rate $\frac{1}{T_1}$ seen in Li NMR \cite{CavaNMR} in the normal state. Scattering rates are comparable to that in metallic Au, which has a 4 to 5 order of magnitude higher conduction electron density. This anomaly might have an origin in a residual superconducting fluctuations that we have proposed.

\section{Discussion}

Bismuth is one of the well studied elemental solids in condensed matter physics. From experimental point of view it continues to surprise us. In the present paper we have studied one such surprise, namely ultra low Tc superconductivity. Focussing on a well known quasi one dimensionality, arising from the 6p valence orbital organization in Bi, we have discussed an interplay of Mott localization, metallization and chain dimerization. We suggest interesting possibilities to arise from even moderate electron-electron repulsion present in 6p orbitals of Bi. 

We have suggested a direction to think about, based on phenomenology and microscopic considerations. A lot remains to be investigated on the heuristics and model we have presented. For example, making use of the hidden one dimensional character one could go beyond RVB meanfield theory and develop coupled chain Bosonization and renormalization theory, including Umklapp terms. This will help one obtain a phase diagram and see how U and interchain hopping compete under renormalization.

One of the consequences of our proposal is presence of low energy spin-1 collective modes supported by fluctuating Mott localization in the chains. It will be interesting to study this issue in detail theoretically and also look for signals in neutron scattering and Raman scattering. Any experiment which tries to bring out hidden Mottness and superconducting fluctuations will be welcome.

We have not explicitly taken into account a strong spin-orbit coupling present in Bi in our theory. Since Kramers theorem replaces real spin by Kramer pair, most of our qualitative conclusion for the bulk are valid. Further the strong covalency (directional bonding in the A7 structure) present in Bi seems to quench the effect of spin-orbit coupling. Situation we encounter is some what similar to Hg, Tl and Pb; they are neighbours of Bi in the same row in the periodic table, having comparable strong spin orbital coupling. Bulk electronic properties in Hg, Tl and Pb, including superconductivity can be discussed by replacing spin by Kramer pair, without invoking spin orbit coupling explicitly. 

The tiny hole and electron Fermi pockets in the BZ are viewed as remnant evanescent Bogoliubov quasi particle excitations in a vacuum containing superconducting fluctuations, after superconductivity has collapsed. Do they leave any direct experimental signatures ?

One of our prediction is a possibility of PT violating order parameter in the recently observed ultra low Tc superconductivity in Bi. It will be interesting to look for this in experiments. 

Surface physics in Bi is an active field now: strong spin-orbit coupling, at the level of band theory, leads to topological phases and phenomena. Surface physics in Bi is likely to become richer from an added dimension of strong correlation effects we have suggested in the present article.

As we mentioned earlier, our theory is applicable to Sb and As, neighbours of Bi in the same column in the periodic table.  Revival of high Tc superconductivity in Bi, As and Sb, through quantum melting of the valence bond crystal seems plausible. Further, theoretical and experimental studies for a better understanding of these systems, with a hope to find high Tc superconductivity and other exotic phases are needed. 

\textbf{Acknowledgement:} I thank S Ramakrishnan (TIFR) for discussion of his results. I thank P.W. Anderson, R.N. Bhatt, N.P. Ong, S. Sondhi and Z. Soos at Princeton for critical remarks; E.H. Lieb for bringing to my attention reference \cite{LiLiebWu}. I am grateful to Science and Engineering Research Board (SERB, India) for a National Fellowship. This work, partly performed at the Perimeter Institute for Theoretical Physics, Waterloo, Canada is supported by the Government of Canada through Industry Canada and by the Province of Ontario through the Ministry of Research and Innovation.

\end{document}